\documentclass{aastex63}

\received{June 1, 2019}
\revised{January 10, 2019}
\accepted{\today}
\submitjournal{AJ}

\shorttitle{Zoo Gems program}
\shortauthors{Keel et al.}

\begin{document}

\title{Gems of the Galaxy Zoos - a Wide-Ranging \emph{Hubble Space Telescope} Gap-Filler 
Program\footnote{This research is based on observations made with the NASA/ESA Hubble Space Telescope obtained from the Space Telescope Science Institute, which is operated by the Association of Universities for Research in Astronomy, Inc., under NASA contract NAS 5-26555. These observations are associated with program 15445.}}

\correspondingauthor{William C. Keel}
\email{wkeel@ua.edu}

\author[0000-0002-6131-9539]{William C. Keel}
\affiliation{Department of Physics and Astronomy, University of Alabama, Box 870324, Tuscaloosa, AL 35487 }

\author{Jean Tate}
\affiliation{Private address\footnote{Deceased 6 November 2020. Jean Tate was among the most active volunteers in
Galaxy Zoo and Radio Galaxy Zoo, and oversaw virtually the entire selection process for Radio Galaxy Zoo targets in this project. Jean set a high standard for the professional scientists on the project, maintaining online material in such good order that
we were able to continue to retrieve and understand even work left in progress.}}

\author[0000-0003-4264-3509]{O. Ivy Wong}
\affiliation{CSIRO Space \& Astronomy, PO Box 1130, Bentley, WA 6102, Australia}
\affiliation{ICRAR-M468, University of Western Australia, Crawley, WA 6009, Australia}

\author[0000-0003-4417-5374]{Julie K. Banfield}
\affiliation{Research School of Astronomy and Astrophysics, Australian National University, Canberra, ACT 2611, Australia}

\author[0000-0001-5578-359X]{Chris J. Lintott}
\affiliation{Department of Physics, University of Oxford, Denys Wilkinson Building, Keble Road, Oxford OX1 3RH, UK}

\author[0000-0003-0846-9578]{Karen L. Masters}
\affiliation{Departments of Physics and Astronomy, Haverford College, 370 Lancaster Ave, Haverford, PA 19041}

\author[0000-0001-5882-3323]{Brooke D. Simmons}
\affiliation{Physics, Lancaster University, Lancaster LA1 4YB, UK}

\author[0000-0002-9136-8876]{Claudia Scarlata}
\affiliation{School of Physics \& Astronomy, University of Minnesota, 116 Church Street S.E., Minneapolis, MN 55455}

\author[0000-0003-4608-6340]{Carolin Cardamone}
\affiliation{Center for the Enhancement of Learning and Teaching, Tufts University, 108 Bromfield Road, Somerville,  MA  02144}

\author[0000-0001-6417-7196]{Rebecca Smethurst}
\affiliation{Department of Physics, University of Oxford, Denys Wilkinson Building, Keble Road, Oxford OX1 3RH, UK}

\author[0000-0002-1067-8558]{Lucy Fortson}
\affiliation{School of Physics \& Astronomy, University of Minnesota, 116 Church Street S.E., Minneapolis, MN 55455}

\author{Jesse Shanahan}
\affiliation{Center for Astrophysics and Space Sciences (CASS), Department of Physics, University of California, San Diego, CA 92093, USA}

\author[0000-0001-8010-8879]{Sandor Kruk}
\affiliation{European Space Agency, ESTEC, Keplerlaan 1, NL-2201 AZ Noordwijk, the Netherlands}
\affiliation{Max-Planck-Institut f\"ur extraterrestrische Physik (MPE), Giessenbachstra\ss e 1, D-85748 Garching bei M\"unchen, Germany"}

\author{Izzy L. Garland}
\affiliation{Physics, Lancaster University, Lancaster LA1 4YB, UK}

\author{Colin Hancock}
\affiliation{Department of Physics and Astronomy, University of Alabama, Box 870324, Tuscaloosa, AL 35487 }

\author{David O'Ryan}
\affiliation{Physics, Lancaster University, Lancaster LA1 4YB, UK}




\begin{abstract}
We describe the Gems of the Galaxy Zoos (Zoo Gems) project, a gap-filler project using short windows in the 
{\it Hubble Space Telescope}'s schedule. As with previous snapshot programs, targets are taken from a pool based on position;
we combine objects selected by volunteers in both the Galaxy Zoo and Radio Galaxy Zoo citizen-science projects.
Zoo Gems uses exposures with the Advanced Camera for Surveys (ACS) to address a broad range of topics in
galaxy morphology, interstellar-medium content, host galaxies of active galactic nuclei, and galaxy evolution. Science cases include 
studying galaxy interactions, backlit dust in galaxies, post-starburst systems, rings and peculiar spiral patterns, outliers from the usual 
color-morphology relation, Green Pea compact starburst systems, double radio sources with spiral host galaxies, and extended 
emission-line regions around active galactic nuclei. For many of these science categories, final selection of targets from a larger list 
used public input via a voting process. Highlights to date include the prevalence of tightly-wound spiral structure in blue, apparently 
early-type galaxies, a nearly complete Einstein ring from a group lens, redder components at lower surface brightness surrounding 
compact Green Pea starbursts, and high-probability examples of spiral galaxies hosting large double radio sources.
\end{abstract}

\keywords{AGN host galaxies (2017) --- Galaxy collisions (585) ---
Starburst galaxies (1570) --- Radio galaxies (1343) --- Ring galaxies (1400)}


\section{Introduction} \label{sec:intro}

Astronomy enjoys a rich history of knowledge gained from
objects at the extremes of sample properties, and outliers to common correlations.
Our experience with spinoff studies from the Galaxy Zoo projects has certainly
borne this out, leading to further observation of rare and unusual
galaxies which in turn yielded insight in a range of questions in galaxy evolution. This paper describes one such project, delivering \emph{Hubble Space Telescope}
(HST) images of galaxies randomly selected from a list chosen for science value in a number of contexts.

Galaxy Zoo has encompassed several iterations of classification based on
volunteer examination of galaxies in digital sky surveys. Initially, ``classic" Galaxy Zoo \citep{GZ2008}
provided broad morphological information (spiral/elliptical/merging, and direction of spiral arms) for
over 900,000 galaxies from data release 7 (DR7, \citealt{dr7}) of the Sloan Digital Sky Survey (SDSS;
\citealt{York}). Galaxy Zoo 2 \citep{GZ2} built on
the demonstrated ability of volunteers to consistently provide finer-grained morphological information,
now working with about 250,000 of the brightest SDSS galaxies. The approach was extended, broadening the
decision tree to encompass clumpy galaxies, to deep optical HST fields in Galaxy Zoo Hubble
\citep{GZH} and the near-IR CANDELS data \citep{Brooke2017}, and most recently images from the Legacy Survey \citep{Dey}. 
The results of these studies led to recognition of the importance of 
blue early-type galaxies (blue ellipticals for short; \citealt{blueE}) and
red spiral galaxies (\citealt{KLMredspirals}, \citealt{Bamford}); \cite{Masters2020} summarizes the first twelve years of Galaxy Zoo results. 
It quickly became clear that the project discussion
forum\footnote{https://www.galaxyzooforum.org/, with content frozen 9 July 2014}, where volunteers could ask questions and exchange comments about galaxy images, was drawing attention
to very rare phenomena, leading to the identification of Green Pea compact starburst systems \citep{Cardamone}, nearly 2000 pairs
of galaxies with overlapping images for dust analysis \citep{overlaps}, and giant extended emission-line regions (EELRs) around active
galactic nuclei (AGN), many of which are so luminous as to suggest that the central AGN must have faded within the
relevant light-travel time (\citealt{voorwerp}, \citealt{voorwerpjes}). Beyond these, numerous other galaxy images of special interest have been brought up for
discussion on the Forum and its successor in the project's Talk interface\footnote{https://talk.galaxyzoo.org/ until April 2019, discussion moved to https://www.zooniverse.org/projects/zookeeper/galaxy-zoo/talk thereafter.}, providing ready sets of objects 
for followup observation. As this became clear, team members could call for specific kinds of objects, and were often 
answered with great energy by volunteers.
Beyond the primary statistical goals of the various iterations of the Galaxy
Zoo public-participation project, some of its hundreds of thousands of 
volunteer participants have identified rare and unusual galaxies 
for which further data would be particularly interesting.

The more recent launch of 
Radio Galaxy Zoo \citep{RGZ}, in which participants examine optical and near-infrared images in concert with
radio data, likewise makes use of the Talk interface for
exchange of more detailed information, particularly on the rare 
radio galaxies with possibly spiral host morphology and on active galactic nuclei (AGN) with
extensive emission-line clouds. For many of the objects, their nature
would become much clearer with higher-resolution optical images than 
the SDSS data used for the initial rounds of Galaxy Zoo classifications, and
numerous specific science goals could be addressed with even
a modest set of such followup images.

Galaxy Zoo team members had long joked about the appropriate followup
observing proposal being ``We have a bunch of weird galaxies, and need 
a closer look to understand them better". This was essentially what the 2017 STScI gap-filler opportunity offered.
We describe in this paper
the resulting program, ``Gems of the Galaxy Zoos" (Zoo Gems for short, program 15445), which has provided 
HST images relevant to a wide range of science cases drawn from Galaxy Zoo and Radio Galaxy Zoo. 
In this paper, we describe these aims, detail how we incorporated public input in selecting the target lists
for many of the science cases we could address, document the setup of the observations, and
present some initial results. While many of the results of Zoo Gems will appear in further papers, we think it useful to 
provide here the common background and rationale of the observations.

\section{Science cases} \label{sec:scicases}

This section sets out the science rationales for various object categories, organized into broad morphological themes. Some categories have had only a single example observed at this point.

\subsection{Galaxy disks}

\subsubsection{Galaxy Zoo: unusual spirals}
This category furnishes a catchall for spiral galaxies - 3-armed grand-design systems, galaxies with very asymmetric patterns but
no obvious interacting companion, or dominant resonance ring structures. The 3-armed spirals took on particular interest with
the finding from Galaxy Zoo 2 classifications that they are not more common in low-density environments, as had been expected from
relative growth properties of various Fourier modes (\citealt{Elmegreen1992}, \citealt{Durbala}), 
and that 
bars are just as common in these as in the 2-armed examples where the 180$^\circ$ symmetry of bar and arms matches 
\citep{Hancock}.

\subsubsection{Galaxy Zoo: nuclear disks and bars in spirals} 

High-resolution images have shown some spiral galaxies, especially those with bars and outer rings, to have 
analogous circumnuclear structures - known as nuclear disks. We included galaxies seen (or
very likely) to have unusually large central disks within bars, or central bars misaligned with the outer structures.

\subsubsection{Galaxy Zoo: backlit galaxies}

Noninteracting galaxy pairs whose images overlap in projection offer a way to study dust attenuation independent of
dust temperature or structure of the galaxy, and subject to completely different systematics than methods relying on
IR emission or modeling of the galaxy's spectral-energy distribution (SED). This approach has been applied 
to very limited sets of galaxies, using ground-based data by  \cite{WhiteKeel92}, \cite{Berlind}, and  \cite{thickthin1},
and with the improved angular resolution of HST imaging by \cite{NGC2207} \cite{thickthin3}, \cite{thickthin4}\ and \cite{angstpair}.
Applicability of this technique remained limited by the very modest number of suitable backlit galaxies then known, a situation
which was dramatically improved by the sensitivity and (especially) dynamic range of SDSS data.
Galaxy Zoo participants provided an extensive finding list of candidate pairs, for a catalog of nearly 2000 such pairs after
validation from their initial examination of DSSS DR7 images alone \citep{overlaps}. This list has been
supplemented by the second pass through DR7 images during Galaxy Zoo 2. Similar pairs were also noted during Hubble Zoo, 
but not considered here since Zoo Gems images would not improve their data quality. The ideal pair would have galaxies
with redshifts so different as to rule out physical association, containing a smooth early-type galaxy behind a relatively symmetric spiral. In
practice these criteria can be relaxed, for example  to include pairs with particular geometries or evidence for very distant attenuation 
regions, as long as the image information is properly used to estimate uncertainties due to departures from symmetry.

\subsection{Starbursts and star-forming regions}

\subsubsection{Galaxy Zoo: Green Peas}

Green Pea systems, as described by \cite{Cardamone}, were initially identified as a class by Galaxy Zoo volunteers. The
name arises from their combination of small size (SDSS Petrosian radius $petrorad\_r < 2.0$\arcsec) and green appearance in
SDSS $gri$ composite images, arising from a strong emission line in the $r$ band (refined to $ugriz$ color criteria by \citealt{Cardamone}). The dominant emission line for these ``green" objects is
redshifted [O III] $\lambda 5007$. Inspection of the SDSS spectra showed most of these to be star-forming systems; their
small angular sizes and redshifts meant these 
are therefore among the most compact star-forming galaxies, and the large emission-line equivalent widths responsible for their 
color selection  mark them firmly as starburst systems. Only a handful of Green Peas appeared
serendipitously in previous HST imagery, so we incorporated Peas into the Zoo Gems list to enable a more systematic study of 
their structures. In particular, we selected filters that emphasized the stellar continuum, for a better view of the structure of
the galaxies themselves, and included systems in three redshift slices. Improved measurements or limits on the sizes of
the stellar structures would lead to better understanding of how intense the starbursts are, and to what extent these 
extreme star-forming regions reside in systems with previous histories of star formation, or show evidence of tidal disturbance which could
trigger these extreme starbursts. These red ACS WFC passbands also provide a contrast with the UV bands previously
observed for some Green Pea samples using HST, which represent the young starburst populations well but not any older stellar 
components.

The Galaxy Zoo sample of Green Peas has generated extensive followup work as sources of Lyman-continuum leakage (\citealt{Yang},
\citealt{Malkan2}), Lyman $\alpha$ emission sources \citep{Orlitova},
extreme starbursts driving global winds (\citealt{Jaskot}, \citealt{Bosch2019}, \citealt{Hogarth}), and
testbeds for chemical evolution scenarios at high star-formation rates (\citealt{Hawley}, \citealt{Amorin}). These compact, intense 
starbursts are likely related to the less massive  ``Little Blue Spheroids" \citep{Moffett}.

\subsubsection{Galaxy Zoo: Poststarbursts}

Spectroscopically-selected post-starburst galaxies\footnote{As it happened, all of the galaxies in 
this input list were affected by an initial bug in SDSS DR7 redshifts, which applied to galaxies having strong
Balmer absorption lines and gave erroneously high redshifts when the gaps between Balmer absorption lines were matched to
broad emission features. This was quickly corrected in the SDSS pipeline, before the release of DR8, and did not affect our sample construction.} (using the combination of H$\delta$ in absorption with equivalent width $> 3 $ \AA\
and H$\alpha$ emission undetected at the $4 \sigma$ level) 
show central concentrations (SDSS \emph{fracdev} parameter) intermediate between disk and early-type galaxies
\citep{Wong2012}. This could reflect genuine morphological transformation or
preferential scales for the starburst, both
issues which higher-resolution images could shed light on. This selection on specific absorption line properties is distinct from the color/luminosity
selection often defining the ``Green Valley", being influenced by star-formation events which are more localized or involve smaller
gas masses, but the possibility of post-starburst systems being seen during morphological
transformation does parallel the inferences about galaxies in the Green Valley largely undergoing a one-time
morphological change (\citealt{Mendez}, \citealt{SchawinskiGV}, \citealt{Smethurst2015}, \citealt{kelvin}).
In addition, the fading starbursts are often concentrated
in knots, which are often well-detected in snapshot observations, allowing broad comparison of the starburst properties.
These properties in combination could suggest to what extent processes in local post-starburst galaxies are, or are not,
useful analogs for the quenching of star formation in the galaxy population more generally. 
\subsubsection{Galaxy Zoo: blue ellipticals}

One of the earliest Galaxy Zoo science results was the existence of galaxies robustly classified as early-type, with colors much bluer than the usual red sequence \citep{blueE}, indicating an unusual level of recent star formation or contamination by the light of an AGN. High-resolution
images can trace the distribution of star-forming regions, indicating whether they form disks, rings, or
the kinds of asymmetric patches that could be infalling star-forming regions.
We tracked three separate subcategories among blue ellipticals (selected using a color cut based on the red sequence track, as in
\citealt{blueE}), so the final list included the highest-ranked examples of blue early-type galaxies with
emission-line ratios from SDSS spectra indicating that AGN or star formation is dominant (5 and 4 targets, respectively), and star-forming 
examples with detected CO or H I emission (3 targets; \citealt{blueECO}, \citealt{BlueEHI}).

\subsubsection{Galaxy Zoo: Red spirals}

Analogous to blue ellipticals, Galaxy Zoo classifications led to identification of a population of red spiral galaxies
\citep{KLMredspirals}, defined by color offset from typical spirals at a given luminosity. While present in all environments, they are 
most abundant in the high-density regions  just outside cluster cores \citep{Bamford}. Galaxies for
voting were selected to be nearly face-on, so the red color is not due solely to attenuation in the disk. The HST images
could show whether the star-forming regions in these systems are of unusually low luminosity or unusually sparse compared
to spirals of typical colors, and address the incidence of bars at small scales for comparison with the global bars which are
common in these systems (\citealt{KLMredspirals11}, \citealt{Kruk2018}). The properties of red spirals may 
give clues to the processes quenching star formation when seen independently of morphological transformations.

\subsubsection{Galaxy Zoo: luminous star-forming clumps in galaxies}

While rare in the local Universe, these may be helpful analogs to the luminous star-forming clumps which are ubiquitous in the
high-redshift galaxy population (\citealt{Cowie1995}, \citealt{Elmegreen2004}). The relative handful of nearby analogous objects will be much better resolved,
providing information on scales of star formation (and sometimes the properties of the most luminous clusters in these
regions).

\subsection{Interacting and merging galaxies}

\subsubsection{Galaxy Zoo: mergers which are very distorted or have very long/luminous tails}

Among the many interacting and merging systems flagged by Galaxy Zoo \citep{Darg}, some stand out even
in such exceptional company as having tidal tails of unusual length or brightness, or main galaxy bodies 
which are unusually distorted. We include some of these in the target list, to sample the properties of galaxy interactions which
produce such extreme stages. Outcomes might include populations of luminous star clusters, morphological information
on scales beyond the SDSS resolution limit, and the role of dust attenuation (which can change our interpretation of 
a system's components and their spatial relationships).

\subsubsection{Galaxy Zoo: collisional and polar rings (including possible lenses)}

Numerous candidates for these rare interaction signatures were noted by Galaxy Zoo participants. They offer
particular insight into not only the prevalence of these kinds of galaxy encounters, but such disparate issues as
the shapes of dark-matter halos, properties of star formation with time mapped to location, and the degree of self-gravitation in polar rings
( \citealt{ReshetnikovCombes}, \citealt{Bizyaev}, \citealt{EgorovMoiseev}). The top-voted polar-ring candidate was earlier catalogued as PRC A-1 in the catalog by  \cite{Whitmore}, supporting the identification of this system as hosting a
classic polar-ring structure. This category also includes objects which may turn out to be gravitationally lensed
arcs when examined at high angular resolution, as indeed happened in one spectacular and near-complete Einstein
ring (section \ref{ring1330}).

\subsubsection{Galaxy Zoo: Red/blue pairs}

Particularly in the earliest examination of Legacy Survey images \citep{Dey}, participants have found a small category of
close pairs of marginally resolved images with very different colors. At first we suspected these might be star/galaxy superpositions, but 
one Zoo Gems image shows at least some to be interesting galaxy interactions, of the general kind
long discussed as mixed-morphology pairs (\citealt{RampazzoSulentic}, \citealt{HernandezToledo}) with contrasting morphologies, colors, and star-forming properties.

\subsubsection{Galaxy Zoo: regrowing disks}

The ``merger hypothesis" for making elliptical galaxies by merging disk systems \citep{Toomre} was largely suggested by the
tendency for disk mergers in simulations to yield elliptical-like, diskless remnants for mass ratios near unity
(in practice, a cutoff near 3:1 has often been taken to give the right fraction of mergers, if not always the right outcomes
in individual cases when such parameters as approach geometry and gas fraction are added to the mass ratio).
However, as merger simulations achieve higher fidelity and are run for more combinations of the numerous
initial conditions, some mergers of near-equal mass galaxies are shown to retain disks afterward (\citealt{bh96},
\citealt{governato}, \citealt{Hopkins2009}).
In parallel, we have noticed a category of interacting-galaxy pairs which show single disks and spiral
patterns surrounding two distinct bulges, which could represent this process in action (and existence of a remnant disk before the
nuclei merge). We included all these systems in the Zoo Gems object list, to give the possibility of empirical information on the survival or
re-formation of disks in major mergers. Such events may be important in producing the population of high-luminosity
``super spiral galaxies" identified by \cite{Ogle2016} and \cite{Ogle2019}. They note that a significant fraction of these
gigantic disk systems show signs of at least minor meters; retaining or regrowing disks after some major mergers
would make it easier to understand the existence of disk-dominated systems at the highest galaxy luminosities.

\subsection{Active galactic nuclei and their host galaxies}

\subsubsection{Galaxy Zoo: EELRs}

Galaxy Zoo participants have proven to be adept at identifying candidates for extended emission-line regions (EELRs)
associated with AGN, based on the distinctive colors in the SDSS \emph{gri} composite images produced by
strong [O III] emission appearing in $g$ or $r$ at different redshifts (\citealt{voorwerp}, \citealt{voorwerpjes}). The 
distance of these clouds from the AGN, and their luminosity compared to what we see from the AGN directly, constrain both
the duty cycle of rapid accretion and characteristic duration of accretion episodes. Fine structure
in these clouds strengthens constraints on the required ionizing luminosity of the associated AGN
(\citealt{HSTvoorwerp}, \citealt{HSTvoorwerpjes}), motivating
us to include the strongest EELR candidates in filters matching the SDSS detection bands, even in the absence of
separate HST continuum data.

\subsubsection{Radio Galaxy Zoo: EELRs}

As in Galaxy Zoo, Radio Galaxy Zoo participants identified many galaxies, or galaxy components, with such extreme colors 
that there are almost certainly strong, resolved emission-line clouds (some tagged as ``RGZ Green" objects, since many are
at redshifts $z \approx 0.25$ where [O III] emission appears in the $r$ band, mapped to
green in the SDSS and analogous color-composite displays). For radio-loud AGN, in addition to photoionization,
emerging jets add the possibility of shock ionization on large scales. Here again, we specified filters
closely matching the detection band from Radio Galaxy Zoo. The input list for voting incorporated both SDSS
colors of the host and spatially resolved structure in the $r$ filter.

\subsubsection{Radio Galaxy Zoo: SDRAGNs}

For convenience, we follow \cite{Leahy} in using DRAGNS (Double-lobed Radio sources Associated with Galactic
Nuclei) to describe typical double-lobed radio galaxies. 
A long-known property of the population of galaxies hosting DRAGNs is that they
are overwhelmingly elliptical galaxies or merger remnants. However, detailed study has revealed a handful of
galaxies, associated at high probability with DRAGNs, with clear spiral structure 
(\citealt{Ledlow}, \citealt{Keel0313},  \citealt{Hota}, \citealt{Mao}, \citealt{Mulcahy})
We will call these rare DRAGNs with spiral host galaxies SDRAGNs.
We concentrate on these because they go against the dominant correlation of luminous double sources with 
elliptical or post-merger host galaxies, offering the possibility of a way to understand which host-galaxy properties
drive (or allow) production of large-scale double sources. The initial list was selected based on location of the
putative host galaxy with respect to the radio lobes, and evidence of a disk
(a large exponential-disk fraction in the SDSS image analysis or from a full 2-dimensional fit
to the SDSS images using GALFIT \citep{GALFIT}).

The angular resolution of HST imaging can make the morphology of these galaxies, many as distant as $z \approx 0.2$, much clearer,
resolving spiral arms, dust lanes, and star-forming regions, allowing a much more confident morphological assessment than the photometric 
bulge-disk decompositions that were our starting points. By the same token, the bulge properties will
also be much better determined, giving an improved understanding of likely black-hole masses and evolutionary paths of these galaxies.
For those radio sources which actually arise in a more distant galaxy than the spiral observed, improved astrometry
with the HST data can identify such false associations.

\subsection{Galaxy Zoo: unusual bulges}

This category includes bulges which are unusually shaped, unusually prominent for the galaxy's morphological type, or
appear prolate with respect to the disk orientation. The latter may indicate that the disk is more like a polar
ring, resulting from material starting in a second galaxy. An important subtype is formed by so-called X-bulges
(extreme versions of ``peanut bulges") which have attracted attention as being edge-on views of bars, allowing us to visualize the 
distribution of stars along a direction which remains unseen in images of more face-on systems (\citealt{BureauFreeman}, \citealt{Kruk2019}).

\section{Final target list}\label{sec:target list}

Our original proposal estimated 1100 targets available. The program was allocated 300 targets for entry into the
whole gap-filler coordinate pool\footnote{One later became impossible to schedule after
tighter restrictions became necessary on guide-star flux.}, so much of the time
available for detailed preparation went into winnowing the target list. Sparse categories (in practice,
those with fewer than 10 examples) were carried along ``as is". 

\begin{figure}
\includegraphics[width=140.mm,angle=270]{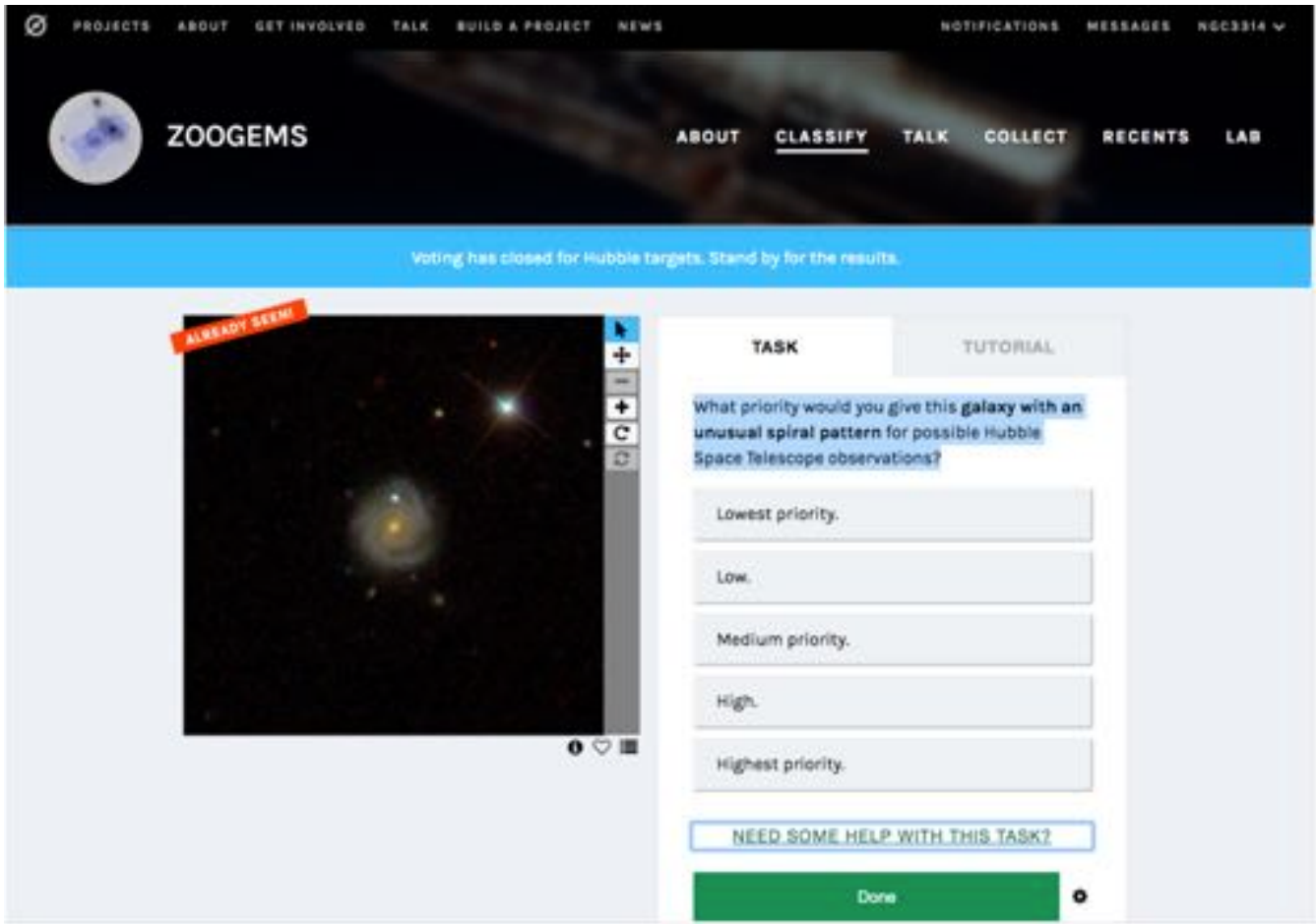}
\caption{Screen capture of the voting interface for Galaxy Zoo targets. Clicking on ``Need some help with this task?" displayed
a longer description and a montage of all galaxies in the same voting category. The red banner at upper left indicates that
a user has already seen this object.
\label{fig-screenshot}}
\end{figure}

Because of the public-participation nature of
the Galaxy Zoo projects, and further encouraged by comments from STScI, we solicited public input to select the
objects to be listed from categories with a large enough number of galaxies. After members of the science team inspected
them for suitability, we set up a web-based voting system, and advertised this opportunity widely through each
project's Web and social-media presence, and coordinated with the STScI social-media team for announcement
through their accounts as well. We used the Zooniverse Project Builder interface\footnote{https://www.zooniverse.org/lab} 
\citep{Trouille}, producing parallel selection
interfaces for the Galaxy Zoo and Radio Galaxy Zoo subsets. The voting was open from 2-18 February 2018,
to meet the 28 February deadline for submission of the Phase 2 proposal with target names, coordinates, and
observation details. The Galaxy Zoo categories drew 6199 votes among 292 galaxies (mean 21.2 per object),
while the Radio Galaxy Zoo objects attracted 9730 votes among 286 galaxies (mean 34.0 each). Numbers of
votes per object varied as users did different numbers per session. The Project Builder interface was set up
so that volunteers would cycle through all members of each science category in sequence, with the option to see a montage
of available objects in that category to allow a more informed comparative ranking.
As illustrated in Fig. \ref{fig-screenshot}, each was presented with a question of the form ``What priority would you give this galaxy with an unusual spiral pattern for possible Hubble Space Telescope observations?" and possible answers 
lowest priority, low, medium priority, high, or highest priority. 

In collating votes, we could distinguish between users who signed in with a Zooniverse account and those
participating anonymously. For the former, registered users, we could recognize multiple votes for a single
object, and counted only the last one. We examined the anonymous votes for any evidence of such misbehavior as
robotic software packing votes, which could affect the outcome. There were very few anonymous votes (180 for Galaxy Zoo,
247 for Radio Galaxy Zoo), widely spread across galaxies, so we saw no evidence of problematic behavior and included
these votes in our rankings. Within each science category being voted on, we ranked objects using a straightforward
mean of votes weighted by priority. Users were given a 5-point scale to select on seeing each galaxy image; we ranked on the mean
with highest priority=1, lowest priority=5. The highest-ranked objects in each science category were carried into the
final target list . (The number of slots allocated for each science category was arbitrarily set by the PI, attempting to
reflect the number of input objects and scientific interest).

The preliminary target list was shared for discussion with the Galaxy Zoo community on the Talk interface. Volunteers identified
some duplicates, because objects could enter in different science categories by different names. We also coordinated
our target list with that of gap-filler program 15446 (Arp and Arp-Madore interacting galaxies, P.I. Julianne Dalcanton), to eliminate duplications while ensuring that some especially interesting systems were on one list or the other.


For Green Peas, whose unresolved SDSS images made visual selection superfluous, the input list consisted of SDSS DR12 objects 
with secure matches to the Portsmouth spectral fitting catalog \citep{Thomas2013}, and BPT type ``Star Forming" from that catalog.  Further selection was for the brightest objects in each of three redshift ranges where ACS filters (mostly) 
isolate continuum. 

\begin{figure}
\includegraphics[width=102.mm,angle=90]{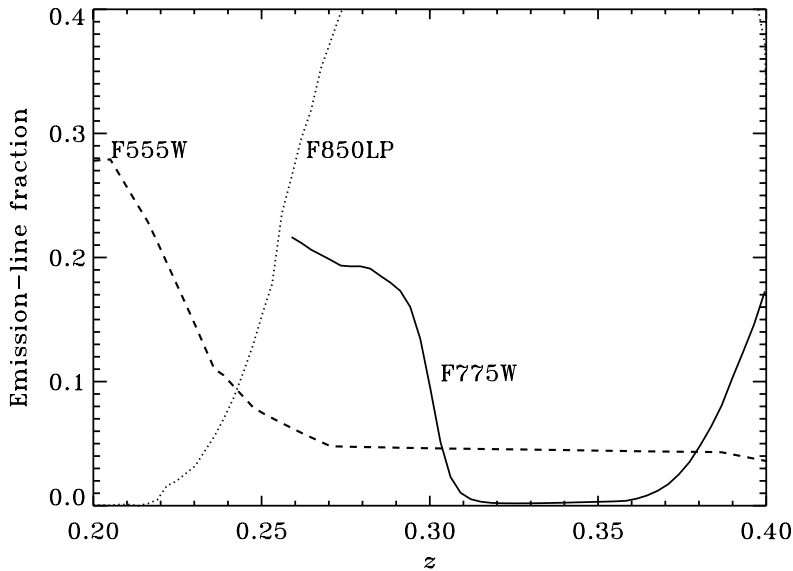}
\caption{Emission-line contribution in the F555W, F775W, and F850LP passbands for a typical Green Pea spectrum, SDSS J131036.73+214817.0 at $z=0.2832$, 
 evaluated at various
redshifts, showing the range near $z=0.24$ where both the very broad F555W and F850LP filters have $<10$\% contribution from emission lines, and the very deep minimum in the F775W filter from $z=0.32-0.36$ where the contamination from line emission
is well below 1\%. \label{fig-GreenPeaF775W}}
\end{figure}

Our category for each targeted object was entered in the phase 2 proposal
file\footnote{Available at https://www.stsci.edu/hst/phase2-public/15445.pdf} under the ``Comments" field.

\begin{figure*}
\includegraphics[width=175.mm,angle=0]{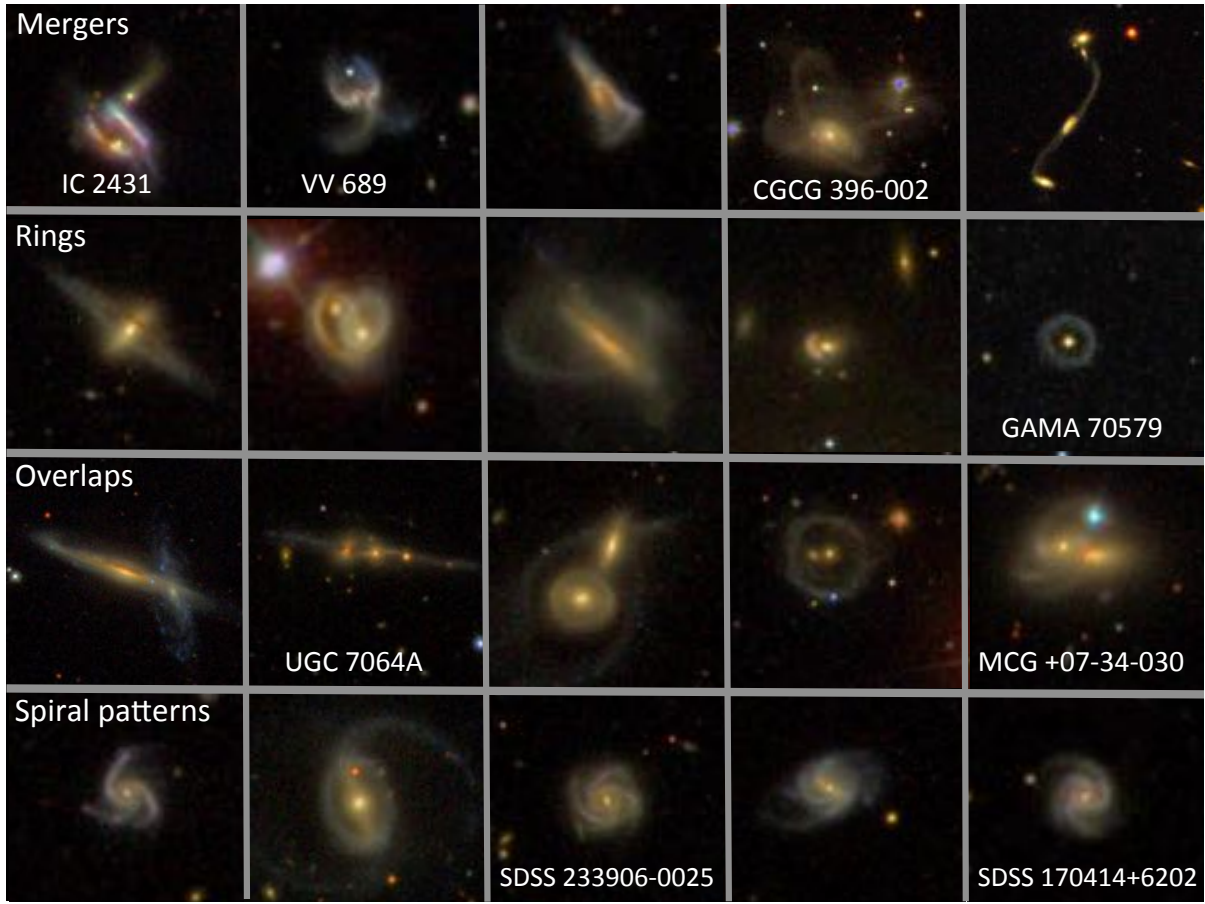}
\caption{Montage of SDSS $gri$ images of galaxies highest-ranked in the voting among four
large categories. Object names are shown for those which have been observed at the time of submission. 
\label{fig-favorites}}
\end{figure*}

Some of the flavor of the voting outcomes can be seen in Fig. \ref{fig-favorites}, showing the top-ranked
galaxies in four categories with numerous objects examined.

Table \ref{tbl-categories} lists the number of target objects in each category, and the number observed as of submission of
this paper. Only primary categories are given, although some fit in multiple categories. For example, all five of the Galaxy Zoo
EELR systems occur in interacting galaxies, and one Radio Galaxy Zoo lens candidate lies in the same ACS field as the Galaxy Zoo EELR candidate SDSS J160646.74+565139.2. For categories where targets were selected by voting, the table also lists the number of objects voted on in each category and the number of votes received. ``Number input" is the number of targets from that category
included in the final observing list, and "Observed" is the number actually observed to date.

\begin{deluxetable}{lcccccl}
\tablecaption{Target and Filter selection by Object Categories} \label{tbl-categories}
\tablehead{
\colhead{Category} & \colhead{Number voted} & \colhead{Votes} & \colhead{Number Input} & \colhead{Observed} &  \colhead{Filter} & \colhead{Goal}    }
\startdata
Galaxy Zoo ring systems  & 26 & 799 &  15 & 6 & F606W & Overall structure  \\
Galaxy Zoo EELRs          & --- & ---  &   5 & 1 & F475W & Line emission\\
Galaxy Zoo regrowing disks   & --- & --- &   7 & 6 & F606W & Overall structure\\
Galaxy Zoo bulges  & --- & --- &   9      & 5 & F814W & Old populations \\
Galaxy Zoo interacting systems & 15 & 515 & 10    & 7 & 606W & Overall structure\\  
Galaxy Zoo red/blue pairs   & --- & --- &  5 & 2 & F606W & Overall structure \\
Galaxy Zoo star-forming clumps & 36 & 904 &  7 & 1 &F606W & Overall structure\\ 
Galaxy Zoo spiral patterns  & 17 & 490 &   7 & 4 & F475W & Young populations\\
Galaxy Zoo poststarbursts   & 56 &  1396  &    20 & 5 & F475W & Young clusters  \\
Galaxy Zoo overlapping pairs    & 63 &  1566 &     21 & 7 & F606W & Dust structure\\
Galaxy Zoo nuclear disks/bars   & --- & ---  &     8 & 4 & F814W & Old populations\\
Galaxy Zoo blue ellipticals     &  61 &  1569 & 12 & 6 & F475W & Spiral structure \\
Galaxy Zoo red spirals     & 19 & 517  &   5 & 1 & F606W & Overall structure\\
Galaxy Zoo Green Peas: all & --- & --- & 74 & 38 & & Stellar continuum \\
Green Peas:	single filter & --- & --- & 47 & 23 & F775W & Stellar continuum \\
Green Peas: 2 filters & --- & --- & 27 & 15& F555W+F850LP & Stellar continuum \\
Radio Galaxy Zoo SDRAGNs & 214 & 7110 & 65 & 32 & F475W & Spiral structure \\
Radio Galaxy Zoo EELRs &  72 & 2620 & 36 & 18 & F625W & Line emission \\
\enddata	 
\end{deluxetable}

\section{Observation setup}\label{sec:observations}

The gap-filler proposal category originated with the realization within STScI that there were schedule gaps too short for typical snapshot 
programs, so that the observational output of HST could be further
enhanced by programs with large target lists, ideally spread around the sky or at least around all right ascensions, which could make
effective use of short observation windows. This rationale, and the review process
for gap-filler programs, are described by \cite{MacKenty}. Use of the Wide-Field Camera (WFC) mode of the 
Advanced Camera for Surveys (ACS; \citealt{ACS}) was
mandated for gap-filler observations, because of its larger field of view than the Wide Field Camera 3 (WFC3),
and to minimize use of a moving mirror in WFC3 identified as a potential failure mode.

An internal STScI pilot project (program 14840) observed bright NGC galaxies, using pairs of 337-second exposures with short dither 
motions in between. All three gap-filler programs finally scheduled use these same exposure times, since the pilot program
had demonstrated that there were a significant number of schedule gaps which 
could accommodate this sequence after guide-star acquisition.

Having settled on the exposure strategy for each object, the available choices were filter,
dither strategy, target location in the ACS field,
and (in a few cases among Green Peas) whether to do 2 dithered exposures or one exposure each in two filters. 
The filter selection followed the science goal - if we were interested mostly in spiral structure, the bluer
F475W (roughly SDSS $g$) filter was chosen to enhance its contrast. If the goal was bulge structure, its signal-to-noise
ratio would be best in the F814W filter. When the goal was emission-line structure already identified in the SDSS $r$ filter, we used the closely-matching F625W filter. The filter choices and rationale for each object category are listed in Table \ref{tbl-categories}.

For Green Pea systems, our aim was to study the continuum structure, using filters dominated by the stellar population rather than 
ionized gas. We were guided by the results of numerically redshifting a typical Pea spectrum and folding through the
system response for several filters (Fig. \ref{fig-GreenPeaF775W}) .
For 35 Green Peas in the redshift range $z=0.32-0.36$, we used the F775W filter, with emission-line fraction $<0.3$\%. For
Green Peas near $z=0.15$ (12 objects), we used F850LP where that is dominated by continuum, while for the 27 
targets near $z=0.25$,
we took single exposures in F555W and F850LP which are each mostly continuum, tolerating the numerous cosmic-ray
events since the targets are small.
Except for this small number of two-filter Green Pea observations, we used a common two-point dither pattern designed to
fill the chip gap (albeit with a central band of cosmic-ray features resulting from coverage with only one exposure in this area).

Target locations on the ACS chips were set in view of the charge-transfer trailing occurring in these CCDs. While
the effects can be substantially reduced with the pixel-based algorithm, essentially a deconvolution, by \cite{AndersonBedin}, the
effect is reduced in the first place if the number of charge transfers can be minimized by placing the target close to the readout amplifier. 
Since orientation had to be unconstrained for snapshot observations, we identified a circular region of interest for each target,
including the SDSS extent of the galaxy and any nearby obvious companions, and used the interactive ALADIN 
viewing feature of the Astronomers' Proposal Tool (APT) to define a POS TARG coordinate offset, so that circular
region closely abutted the edge of CCD WFC1. We considered putting targets closer to the corner of the overall
ACS WFC field where optical distortion gives the largest pixels on the sky, in order to improve surface-brightness
sensitivity, but that region falls on CCD WFC2 whose slightly higher readout noise more than compensates for the
pixel-area difference in sensitivity.

During this stage, we also caught some pasting errors in the sign of target declinations, possibly fostered by
having target lists and formats from multiple sources.

\section{Sample results}\label{sec:results}

The Zoo Gems observations to date are listed in Table \ref{tbl-obs}, by science topic and observation date. As of 31 January 2022, 146 objects had been observed, 49\% of the input list. The total
exposure time in each case is 674 seconds, in two dithered exposures when a single filter was used, or two individual 337-second
exposure for those Green Peas with two filters listed. For these 2-filter objects, we list the final two characters of the second data set identifier after the complete identifier for the first exposure. In a few cases, one exposure was terminated onboard before the
planned duration.

The ACS images showed some objects to be morphologically rather different than we anticipated from SDSS images.
In these cases, we show them in Table  \ref{tbl-obs} and in our further discussion according to the category where they fit most 
closely rather than the category listed in the proposal. One object, SDSS J222024.58+010931.3, turned out to be a superposition of a star and faint background galaxies which
mimicked a ring structure in the SDSS data, and is not included in Table \ref{tbl-obs}. 

The following sections highlight some initial results from the Zoo Gems observations, and demonstrate the value of even
such shallow exposures in addressing a variety of scientific questions.


\startlongtable
\begin{deluxetable}{llcccl}
\tablecaption{Zoo Gems Observations}
\tablehead{
\colhead{Dataset} & 	\colhead{Target Name} & \colhead{ $\alpha_{2000}$} & \colhead{$\delta_{2000}$} & \colhead{UT Start Time} & \colhead{Filter} }
\startdata
\multispan2  {\bf Unusual spiral patterns:} \hfill \hfill & \\
JDS452010 & SDSS-170414.33+620234.0	& 17 04 16.225	& +62 02 57.91	& 2018-07-06 08:19:20		& F475W 	 	\\ 
JDS450010 & UGC-4250					& 08 10 05.486	 & +46 11 34.90 & 2021-01-11 05:23:05	         & F475W \\
JDS451010 & NGC-2595			                & 08 27 42.024	& +21 28 44.76	& 2021-03-16 11:57:55             & F475W\\
JDS453010 & MCG+10-21-019 & 14 36 35.851 & 	+57 47 49.08	& 2021-08-03 20:07:36	                        & F475W \\
JDS454010 & SDSS-233906.23-002615.0        & 23 39 06.236 &  -00 26 15.10 & 2021-11-13 12:36:39          & F475W\\
{\bf X-Bulges:} & \\
JDS439010 & NGC-1175					& 03 04 30.752	& +42 20 07.55	& 2019-07-18 02:34:29		& F814W	\\
JDS436010 & SDSS-1237661418212229211	& 14 09 04.333	& +54 52 19.88	& 2019-09-07 11:43:00	 	& F814W \\	
{\bf Nuclear disks/bars:} & \\
JDS498010 & CGCG-245-033	                        & 13 12 56.707	& +47 27 23.84	& 2021-03-12 22:49:26	 & F814W\\	
JDS40A010 & 	NGC-2771				& 09 10 41.192	& +50 22 36.54	& 2021-03-14 01:09:33	& F814W\\
JDS40B010 &	NGC-5945				& 15 29 45.007 & +42 55 07.13 & 2021-03-14 04:40:37	& F814W\\
JDS40F010 & MCG+09-19-030				& 11 17 43.510	& +53 47 36.25	& 2021-09-27 04:08:05	& F814W\\
\multispan2 {\bf Large or prolate bulges:} \hfill \hfill & \\
JDS420010	& NGC-0810				& 02 05 28.560	& +13 15 05.76	& 2019-06-24 18:54:23		& F606W	\\
JDS435010	& CGCG-308-012			& 06 13 05.304	& +64 33 41.04	& 2019-12-19 07:05:33		& F814W	\\
JDS47T010	& UGC-10374			        &16 23 39.166	& +50 58 09.96	& 2020-03-03 13:11:57		& F814W	\\
{\bf Red/Blue Pairs:} & \\
JDS432010	& SDSS-1237678618479755694	& 22 12 06.397	& +01 46 06.37	& 2019-11-28 19:29:10		& F606W	\\
JDS431010	& GAMA-636444		 & 09 23 21.888 & -01 43 33.96 & 2022-01-29 18:48:26 & F606W\\
{\bf Star-forming knots:} & \\
JDS430010	& GAMA-302130			& 09 05 46.392	& +01 15 32.76	& 2021-10-06 04:06:12		& F606W\\
{\bf Overlapping Galaxies:} & \\
JDS437010	& SDSS-1237661417676800323	& 14 28 14.096	& +53 14 29.00	& 2019-05-08 08:33:55	& F814W	\\
JDS490010	& SDSS-115331.86+360024.2		& 11 53 31.865	& +36 00 24.27	& 2019-05-28 18:56:44		& F606W	\\
JDS478010	& UGC-7064A					& 12 04 44.973	& +60 40 24.56	& 2019-07-12 11:33:01		& F606W	\\
JDS488010	& NGC-5021					& 13 12 06.265	 & +46 11 45.75 & 2020-07-16 12:05:54 &  F814W \\
JDS489010	& UGC-12281					& 22 59 14.839	 & +13 36 16.74 & 2020-09-01 04:48:39 & F606W \\ 	 
JDS485010	& IC-720				  		& 11 42 22.336	& +08 46 11.51	& 2021-03-14 06:10:29	& F606W\\
JDS481010	& MCG+07-34-030				& 16 25 58.133	& +43 57 46.47	 & 2021-10-10 11:32:38		& F606W\\
\multispan2 {\bf Interacting/Merging Systems:} \hfill \hfill & \\
JDS449010 	& SDSS-081913.94+591926.4		& 08 19 13.920	& +59 19 26.76	& 2019-02-22 21:23:13		& F606W \\
JDS47W010	& UGC-00240					& 00 25 10.106	& +06 29 27.17	& 2019-10-05 06:19:39		& F606W	\\
JDS44M010	& VII-ZW-090					& 10 36 35.625	& +02 21 31.41	& 2020-02-27 03:33:10		& F475W \\
JDS428010	& SDSS-095346.77-012746.1		& 09 53 46.680	& -01 27 45.00	& 2020-03-03 05:50:45		& F606W	 \\
JDS406010	& SDSS-1237668504364187727	& 16 12 24.606	& +59 46 10.47	& 2021-03-14 03:53:56            & F475W \\
JDS444010	& CGCG-396-002			       & 05 37 35.976	& +01 20 04.20	& 2021-03-18 22:36:26		& F606W\\
JDS442010	& VV-689	                                          & 10 01 39.502	& +19 47 32.58	& 2021-04-30 17:06:39	        & F606W \\
JDS441010	& IC-2431						& 09 04 34.776	& +14 35 45.96	& 2021-10-03 02:54:35		&F606W\\
\multispan2  {\bf Ring(ed) Galaxies, Lenses:} \hfill \hfill & \\
JDS425010	& SDSS-1237679438812676365	& 02 03 28.727	& -06 59 49.72	& 2019-02-17 01:28:33		& F606W \\
JDS426010	& CGCG-087-009				& 07 33 17.712	& +18 17 24.36	& 2019-03-20 14:17:24		& F606W\\\
JDS418010	& GAMA70579					& 12 01 43.464	& +00 10 59.16	& 2019-05-17 06:34:36		& F606W\\\
JDS424010	& SDSS-133145.32+513431.2		& 13 31 45.326	& +51 34 31.22	& 2019-09-08 17:53:10		& F814W	\\	
JDS495010	& IC-3828						& 12 50 20.695	 & +37 56 56.19 & 2020-06-13 17:27:34 &  F606W \\ 
JDS405010	& SDSS-1237678595932094536	& 22 20 24.589	& +01 09 31.30	0 & 2020-12-16 21:52:36 &	F475W\\
JDS427010	& SDSS-081740.08+042952.3		& 08 17 40.080	& +04 29 52.44	0 & 2021-01-12 21:07:04	& 	F606W\\ 
\multispan2  {\bf Galaxy Zoo EELRs:} \hfill \hfill  & \\
JDS403010	& SDSSJ160646.74+565139.2		& 16 06 46.740	& +56 51 39.20	& 2021-12-24 16:13:14	& F606W\\
\multispan2  {\bf Radio Galaxy Zoo EELRs:} \hfill \hfill  & \\
JDS46N010	& SDSS-130854.52+562155.6		& 13 08 52.460	& +56 22 42.40	& 2019-02-15 02:12:01		& F625W \\
JDS46Z010	& SDSS-075529.95+520450.6		& 07 55 26.309	 & +52 03 51.25 & 2019-04-02 07:46:09		& F625W	\\ 	 
JDS46U010	& SDSS-102733.29+544227.9		& 10 27 30.688	& +54 41 30.79	& 2019-06-04 03:47:14		& F625W	\\
JDS46R010	& SDSS-121849.88+502617.6		& 12 18 45.908	& +50 25 30.87	& 2019-06-07 00:40:21		& F625W	\\
JDS46E010	& SDSS-160344.95+524220.6		& 16 03 42.715	 & +52 41 25.34 & 2019-08-20 14:58:52		& F625W	\\
JDS47M010	& SDSS-010206.98+093427.6		& 01 02 03.668	& +09 35 04.12	& 2019-10-08 21:44:09		& F625W	\\
JDS47G010	& PKS-0236+02				& 02 38 34.584	& +02 34 46.84	& 2019-10-16 03:00:46		& F625W	\\
JDS47V010	& SDSS-101147.31+071915.2		& 10 11 50.739	& +07 19 42.52	& 2020-02-15 07:07:13		& F625W \\
JDS46H010	& SDSS-025210.17+025430.1		& 02 52 10.017	& +02 53 32.90	& 2020-03-06 11:15:15		& F625W \\
JDS47C010	& SDSS-141119.04+094225.3	        & 14 11 19.04    & +09 42 25.3 & 2020-08-02 17:12:10		& F625W \\
JDS46I010	& SDSS-105426.23+573649.1		& 10 54 32.316	 & +57 37 16.16 & 2020-10-17 09:26:49		& F625W \\
JDS47F010	& SDSS-082400.50+031749.4		& 08 24 04.353 &  +03 18 07.73 & 2021-01-14 06:29:16 		& F625W\\	
JDS46C010	& B2-0832+34					& 08 35 15.391	& +34 34 01.90	& 2021-03-12 06:17:49	       	& F625W\\
JDS46W010	& SDSS-083512.43+175441.0		& 08 35 10.668	& +17 53 50.09	& 2021-03-19 01:55:27	        & F625W\\
JDS47D010	& SDSS-123300.30+060326.1		& 12 32 58.539	& +06 04 16.38	& 2021-03-21 08:13:26		& F625W\\ 
JDS46P010	& 3C-458				                 & 23 12 48.445	& +05 17 05.36	& 2021-09-10 03:03:35		& F625W\\
JDS46Q010	& 4C+08.70					& 23 36 40.401	& +08 49 55.09	& 2021-09-12 04:19:16 		& F625W\\
JDS46O010	& SDSS-141408.44+484156.0		& 14 14 08.445  & +48 41 56.00 & 2021-09-22 01:59:54		& F625W\\
\multispan2  {\bf Blue Elliptical Galaxies:} \hfill \hfill & \\
JDS40O010	& CGCG-315-014				& 12 06 17.055	& +63 38 19.08	& 2019-04-15 14:34:09		& F475W	\\
JDS40M010	& MKN-0888					& 16 44 30.755	& +19 56 26.73	& 2019-08-07 09:15:53		& F475W	\\
JDS40P010	& SDSS-111850.04+422541.8		& 11 18 50.047	& +42 25 41.84	& 2019-12-28 23:11:30		& F475W	\\
JDS40H010	& SDSS-031749.30+011337.1		& 03 17 49.304 & +01 13 37.25 & 2020-11-28 06:10:32             & F475W \\
JDS40J010	& SDSS-000907.90+142755.8	        & 00 09 07.908	& +14 27 55.83	 & 2021-08-06 09:06:11		& F475W \\
JDS40G010	& CGCG-432-030				& 23 47 03.791 & +14 50 30.36	 & 2021-09-17 06:36:51		& F475W\\
{\bf Red spiral galaxies:} & \\
JDS40T010	& UGC 3935					& 07 37 49.410	 & +46 23 51.53 & 	2020-10-18 00:25:12		& F606W\\
\multispan2  {\bf Regrowing-disk Mergers:} \hfill \hfill & \\
JDS414010	& NGC-2292					& 06 47 40.830	& -26 45 05.00	& 2020-01-23 04:22:48		& F606W	\\
JDS412010	& CFHTLS1220-215555			& 08 50 58.169	& -04 02 12.85	& 2020-03-12 02:41:03		& F606W \\
JDS410010	& UGC-4052					& 07 51 16.564	& +50 14 03.27	& 2020-05-01 11:33:05		& F606W \\
JDS408010	& SDSS-1237659936978568047	& 00 43 41.784	 & +43 02 35.16 & 2020-12-19 19:47:02 		& F606W\\
JDS413010	& SDSS-1237680246274064522	& 23 26 23.853	& +19 27 09.12	& 2021-09-20 01:17:10		& F606W\\
JDS409010	& SDSS-1237678439701807265	& 02 49 03.312	& +03 12 12.60	& 2021-09-21 09:16:29		& F606W\\
\multispan2  {\bf Post-Starburst Galaxies:} \hfill \hfill & \\
JDS464010	& CGCG-292-024				& 11 44 52.092	& +57 52 24.67	& 2018-08-09 03:21:12	& F475W	  \\
JDS474010	& SDSS-124354.11+163250.5		& 12 43 54.178	& +16 32 50.85	& 2019-03-19 13:30:53 	& F475W \\
JDS460010	& NGC-3156					& 10 12 41.183	& +03 08 04.71	& 2020-02-24 16:44:52	& F475W	\\
JDS457010	& UGCA-188			                & 09 55 29.700	& +08 23 26.28	& 2020-06-12 20:44:08	& F475W \\
JDS476010	& VCC-1711			                 & 12 37 22.147	& +12 17 13.32	& 2021-03-21 19:20:48	& F475W\\
{\bf SDRAGNs:} & \\
JDS45H010	& SDSS-091949.07+135910.7		& 09 19 47.195	& +13 58 22.68	& 2018-05-15 21:49:55	& F475W 	 	\\ 
JDS43Y010	& UGC-1797					& 02 19 58.728	& +01 55 48.72	& 2018-07-03 02:46:37	& F475W 	 \\
JDS44C010	& SDSS-16562058+6407529		& 16 56 16.945	& +64 07 14.62	& 2018-08-24 13:52:36	& F475W	 \\	 
JDS45T010	& SDSS-112811.63+241746.9		& 11 28 09.853	& +24 18 39.94	& 2019-02-23 05:51:12	& F475W \\
JDS45Z010	& B3-1352+471					& 13 54 30.924	& +46 56 44.51	& 2019-04-28 00:35:49	& F475W	\\
JDS44X010	& SDSS-132809.31+571023.3		& 13 28 03.443	& +57 10 13.25	& 2019-05-14 07:06:26	& F475W	\\
JDS45J010	& SDSS-163300.85+084736.4		& 16 32 58.024	& +08 47 03.84	& 2019-07-11 23:11:15	& F475W	\\
JDS44Z010	& B2-1644+38					& 16 46 25.987	& +38 31 03.31	& 2019-07-19 17:39:53	& F475W	\\
JDS43V010	& SDSS-172107.89+262432.1		& 17 21 05.558	& +26 23 54.34	& 2019-08-22 16:18:28	& F475W	\\
JDS47J010	& SDSS-134900.13+454256.5		& 13 49 06.051	& +45 43 03.52	& 2019-11-13 21:02:54	& F475W	\\
JDS45V010	& SDSS-214110.61+082132.6		& 21 41 11.564	 & +08 20 35.91 & 2019-12-11 14:04:35	& F475W	\\
JDS45G010	& SDSS-081303.10+552050.7		& 08 13 00.417	& +55 21 37.26	& 2019-12-25 01:19:10 	& F475W	\\ 
JDS44D010	& SDSS-150903.21+515247.9		& 15 09 08.415	& +51 53 28.37	& 2020-01-14 13:19:06 	& F475W	\\
JDS44R010	& B2-0938+31A				& 09 40 59.773	& +31 26 29.12	& 2020-02-13 02:39:49	& F475W \\
JDS47H010	& SDSS-095605.87+162829.9		& 09 56 01.712	& +16 28 52.37	& 2020-02-14 21:34:14	& F475W \\
JDS45A010	& IC-4234						& 13 22 58.535	& +27 07 09.45	& 2020-04-09 03:03:16	& F475W \\
JDS44J010	& SDSS-080658.46+062453.4	        & 08 06 58.46 & +06 24 53.4 & 2020-05-28 23:07:53	 & 	F475W \\
JDS44P010      & SDSS-095833.44+561937.8		& 09 58 39.333	 & +56 20 16.06 & 2020-10-17 13:19:01	& F475W\\
JDS44I010	& SDSS-080259.73+115709.7		& 08 03 04.152	& +11 57 33.22	 & 2021-01-10 03:57:52	 & F475W\\
JDS47K010	& SDSS-113648.57+125239.7		& 11 36 48.57 	& +12 52 39.7	& 2021-03-14 02:58:57	& F475W\\
JDS41L010	& B3-0852+422			                 & 08 55 44.151	& +42 03 44.70	& 2021-03-14 23:24:25	& F475W\\
JDS45L010	& SDSS-090305.84+432820.4		& 09 03 02.290	& +43 27 51.56	& 2021-03-15 00:59:46	& F475W\\
JDS44G010	& SDSS-082312.91+033301.3		& 08 23 11.663	& +03 32 03.79 & 2021-03-15 05:49:08	& F475W\\
JDS45B010	& SDSS-083351.28+045745.4		& 08 33 50.118	& +04 56 54.67	& 2021-03-18 00:32:22	& F475W\\
JDS45F010	& SDSS-163624.97+243230.8		& 16 36 21.753	& +24 32 46.26	& 2021-06-05 19:19:04	& F475W\\
JDS44K010	& SDSS-083224.82+184855.4		& 08 32 27.328	& +18 49 54.04	& 2021-09-28 05:30:40	& F475W\\
JDS45E010	& SDSS-130300.80+511954.7		& 13 03 00.803	& +51 19 54.70	& 2021-10-01 05:04:08	 & F475W\\
JDS45I010	& SDSS-084759.90+124159.3		& 08 47 59.90 	& +12 41 59.3	& 2021-10-02 04:44:49	 & F475W\\
JDS44T010	& SDSS-020904.75+075004.5		& 02 09 04.750	& +07 50 04.50	& 2021-10-22 03:54:22	& F475W\\
JDS45W010 	& SDSS-090147.17+164851.3          & 09 01 47.17   & +16 48 51.3 & 2021-11-13 10:05:19	& F475W\\
JDS43Z010	& B3-0911+418			                & 09 14 45.528	& +41 37 14.52	& 2021-12-29 08:17:30  	& F475W\\
JDS47L010	& SDSS-092605.17+465233.9		& 09 26 05.17	& +46 52 33.9	& 2021-12-29 19:23:05     & F475W\\
{\bf Green Peas:} & \\
JDS42MO4Q/5Q & SDSS-1237651537646977181	& 15 04 57.987	& +59 54 07.27 & 2018-07-10 14:11:15	& F555W, F850LP	 \\
JDS42KCEQ/FQ & SDSS-1237659330316140926	& 16 33 37.941	& +37 53 14.30	& 2018-07-14 03:38:45	& F555W, F850LP	 \\
JDS41D010	& SDSS-1237648720155902209	& 11 49 46.471	& -01 02 17.65	& 2018-11-29 21:59:41	& F775W	 \\
JDS41F010	& SDSS-1237661386530882021	& 14 10 05.248	& +53 50 37.89	& 2018-12-27 01:56:41	& F775W	 \\	 
JDS40X010	& SDSS-1237651271358349509	& 10 26 15.207	& +63 33 08.49	& 2019-01-02 02:51:41	& F775W	 \\	 
JDS40Y010	& SDSS-1237657878078751164	& 08 08 16.907	& +28 14 31.14	& 2019-02-24 00:38:23	& F775W \\
JDS42IUGQ/HQ & SDSS-1237654398623023110	& 13 36 07.914	& +62 55 30.77	& 2019-04-12 02:35:01	& F555W, F850LP	\\
JDS40Z010	& SDSS-1237651271899939173	& 12 05 17.538	& +66 40 29.64	& 2019-04-19 05:30:54	& F775W	\\
JDS42ZV5Q/6Q & SDSS-1237667551414452597 	& 10 15 41.152	& +22 27 27.52	& 2019-05-16 16:09:08	& F555W, F850LP	\\
JDS42QJPQ/QQ & SDSS-1237661851469021323	& 12 14 23.180	& +45 20 40.91	& 2019-07-17 04:52:31	& F555W, F850LP	\\
JDS42B010	& SDSS-1237669698364768508	& 21 08 03.059	& +05 27 07.14	& 2019-08-10 04:15:11	& F775W	\\
JDS41A010	& SDSS-1237651273498755389	& 08 38 40.165	& +54 44 03.49	& 2019-10-01 17:20:01	& F775W	\\
JDS43DUMQ/NQ & SDSS-1237666300559098162	& 03 53 32.464	& -00 10 28.88	& 2019-11-14 06:16:30	& F555W, F850LP \\
JDS42F010	& SDSS-1237680507722793618	& 23 19 27.467	& +33 23 24.76	& 2019-11-25 15:15:51	& F775W	\\
JDS43EDKQ/LQ & SDSS-1237665369575981438	& 10 20 57.462	& +29 37 26.47	& 2020-01-26 23:25:54	& F555W, F850LP	\\
JDS42WWCQ/DQ & SDSS-1237667536933945523	& 10 04 00.641	& +20 17 19.25	& 2020-02-21 03:03:38	& F555W, F850LP \\
JDS43T010	& SDSS-1237664668421849521	& 08 15 52.002	& +21 56 23.65	& 2020-04-27 20:20:00	& F850LP	 \\
JDS42LSNQ/OQ	& SDSS-1237658491735507237	& 10 55 30.41    & +08 41 32.8  & 2020-06-03 14:22:26	& F555W, F850LP \\
JDS43M010	& SDSS-1237657632187613477	& 09 24 38.718	& +47 07 58.93	& 2020-06-10 20:58:36	& F850LP \\
JDS41Q010	& SDSS-1237662300818702524	& 13 01 28.316	& +51 04 51.18	& 2020-08-01 18:46:22	& F775W \\
JDS42NVWQ/XQ	& SDSS-1237653653450064110 & 00 42 36.92 & +16 02 02.7	& 2020-11-12 16:48:32	& F555W, F850LP \\
JDS42HVJQ/KQ	& SDSS-1237660343930782057 & 08 45 11.669 & +32 51 53.92 & 2020-12-09 06:05:51	& F555W, F850LP \\
JDS41G010		& SDSS-1237654383585395027 & 08 34 40.056 & +48 05 40.91 & 	2021-01-11  21:16:54 & 	F775W\\
JDS42E010	& SDSS-1237667254538142046	& 09 51 03.165	 & +24 54 35.70 & 2021-02-07 10:29:06	& F775W\\
JDS42XQMQ/NQ	& SDSS-1237667211050680599 & 09 41 49.637 & +23 37 30.07 & 2021-02-14 12:28:58 & F555W, F850LP \\
JDS41U010	& SDSS-1237662640662905610	& 16 46 12.15   &   +20 54 11.5 & 2021-03-13 05:34:25 & F775W\\
JDS43J010	& SDSS-1237658423543529721	& 09 05 35.161	& +04 53 34.51 & 2021-03-18 02:09:30 & F850LP\\
JDS42SAJQ/KQ & SDSS-1237648704041976118	& 12 29 33.142	& -00 18 01.68	& 2021-03-26 05:48:14 & F555W, F850LP\\
JDS42UHCQ/DQ & SDSS-1237667781210145026	& 10 04 34.733	& +17 47 35.35	& 2021-04-27 00:09:42 & F555W, F850LP\\
JDS43CVFQ/GQ & SDSS-1237679476939882858	& 01 03 21.059	 & +21 32 15.91 & 2021-09-09 17:36:16 & F555W, F850LP \\
JDS43N010	& SDSS-1237655373039927410	& 16 53 04.490	& +33 39 37.74	& 2021-09-09 18:57:15	& F850LP \\
JDS43O010	& SDSS-1237663782590021909	& 00 29 38.169	& -01 12 16.05	& 2021-09-17 05:07:19	 & F850LP\\
JDS41R010	& SDSS-1237657589239382254	& 09 54 22.599	 & +47 51 44.02 & 2021-09-22 03:22:49	& F775W\\
JDS41S010	& SDSS-1237661382770098472	& 09 20 36.046	 & +32 42 52.63 & 2021-09-29 03:39:51	 & F775W\\
JDS41J010	& SDSS-1237663204918952341	& 00 44 00.266	& +00 47 24.68	& 2021-09-30 23:56:42	& F775W\\
JDS41I010	& SDSS-1237657628456386802	& 12 33 38.626	& +51 41 59.34	& 2021-10-01 01:52:15	& F775W\\
JDS42JI3Q/4Q	& SDSS-1237660635458568341	& 10 27 16.725	& +43 42 02.18	& 2021-10-03 06:30:17	& F555W,F850LP\\
JDS41P010	& SDSS-1237661361304568164	& 14 46 42.608	& +40 48 44.21	& 2021-12-11 16:35:48	& F775W\\
\enddata	 
\label{tbl-obs}
\end{deluxetable}

\subsection{Green Peas}

\begin{figure*}
\includegraphics[width=175.mm,angle=0]{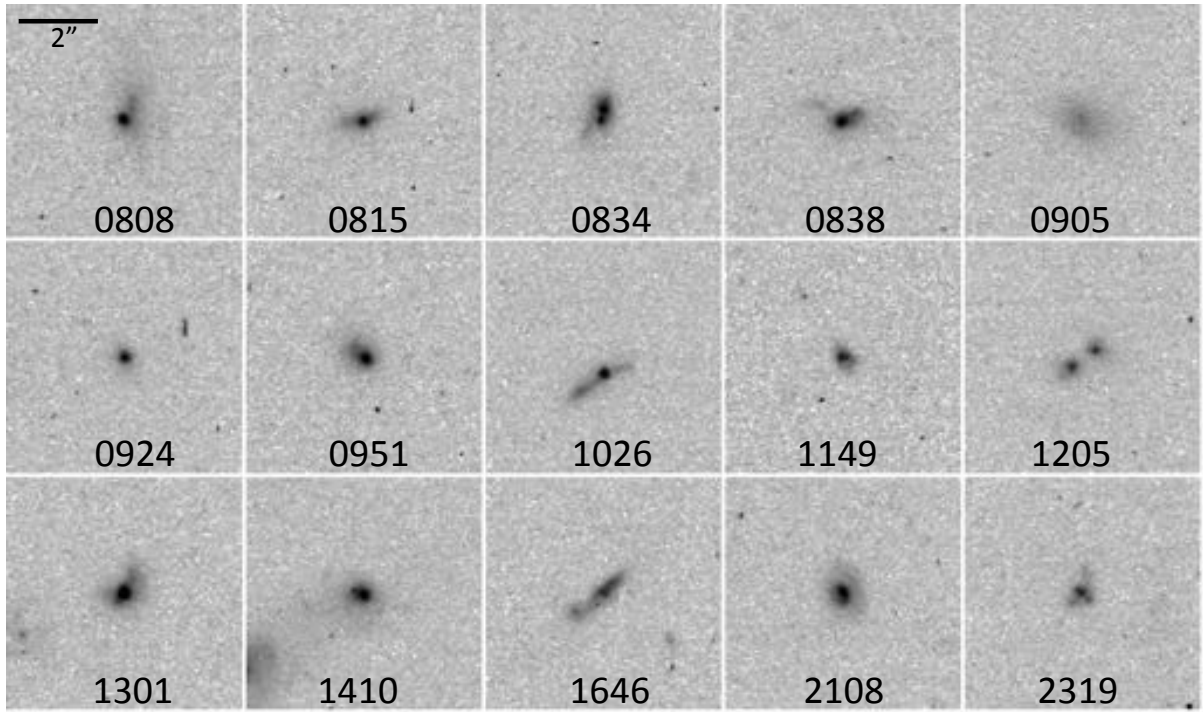}
\caption{ The first 15 Green Pea systems observed in single filters, as listed in Table \ref{tbl-obs}. Abbreviated names showing the
first 4 digits of the right ascension are used for convenience. Each panel is a $6 \times 6$\arcsec
region, typically $24 \times 24$ kpc, and all are 
shown to the same intensity scale, logarithmic above a slight negative offset. North is up, and east to the left in 
each case. The 2\arcsec scale bar matches the limiting Petrosian radius from SDSS data used in the sample selection. 
All were observed in the F775M band except 0815, 0905, and 0924, observed in F850LP. 
\label{fig-gpmontage}}
\end{figure*}

The great majority of Green Peas (34/38, 89\%) are resolved into multiple distinct components or show surrounding non-axisymmetric structure. The structures include double components, tails, and apparent disks. 
Four systems (SDSS J004236.92+160202.7, SDSS J092438.71+470758.9, SDSS J100400.64+201719.2, and SDSS J165304.48+333937.7) have central peaks only
marginally resolved in the ACS data; a simple Gaussian comparison with star images suggests intrinsic FWHM $< 1.8$ pixels (0.09\arcsec ) or 0.36 kpc at the typical $z=0.25$.
Some of the Green Pea systems show tidal tails
or patchy spiral patterns; most have either multiple knots or show well-resolved surrounding galaxies (shown in Fig. \ref{fig-gpmontage}
for those observed in a single filter, allowing easy rejection of cosmic-ray artifacts). Analysis of the 2-filter 
Zoo Gems data by \cite{leonardo} indicates that the intense starburst regions are surrounded by redder 
components, most likely older stellar populations. The smallest sizes we measure in these deep-red bands are 
comparable to the UV sizes measured (predominantly for the brightest components) by \cite{Yang} and \cite{Kim}.

\subsection{Blue ellipticals}

\begin{figure*}
\includegraphics[width=175.mm,angle=0]{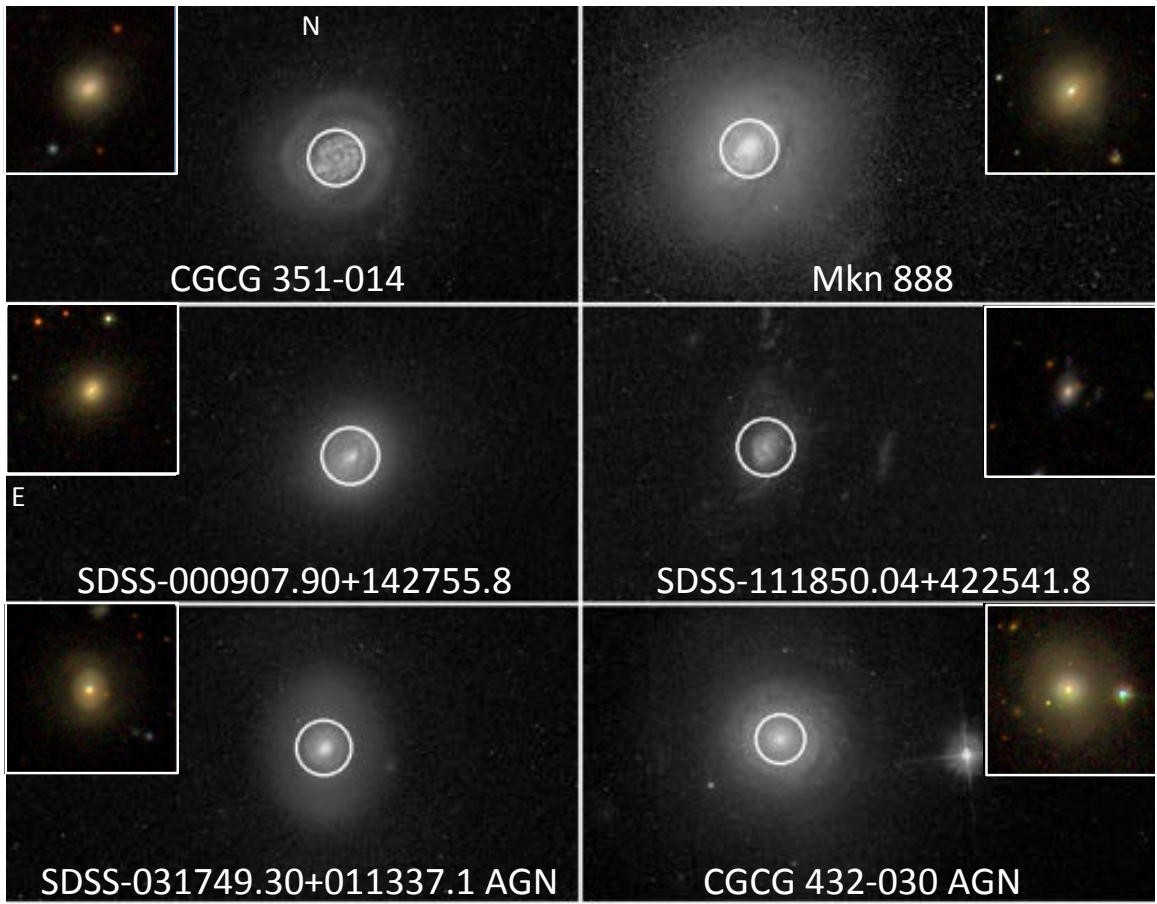}
\caption{Blue early-type galaxies, in the F475W filter to emphasize young stellar populations. The bottom two are classified as having
spectroscopic AGN by the SDSS automated system. Each galaxy shows a tightly wound spiral pattern
near the nucleus; the superimposed circles have radius 2.5\arcsec, showing how strongly these patterns are blended together in typical
survey images. Insets show SDSS composite images, 60\arcsec  square, as used by Galaxy Zoo participants in the initial
classifications. Some show various levels of structure in these images, but only CGCG 432-030 might have been classified as a clear spiral from SDSS data. North is up, and east to the left in each case.
\label{fig-bluee}}
\end{figure*}

Each of the six blue early-type galaxies observed in Zoo Gems shows a distinct spiral structure, which in each case is
too tightly wound to have been resolved in SDSS data (Fig. \ref{fig-bluee}). These are not simply otherwise-normal elliptical galaxies
with scattered star-forming regions, although in some cases the outer light distribution is less disc-like than the inner
regions. In fact, two of these galaxies, Mkn 888 and SDSS J031749.30+011337, have a nearly pure $r^{1/4}$ profile as assessed 
in the SDSS {\it fracDeV} parameter, with values 0.97--1 among all SDSS filters. The other four have values 0.49--0.96 in
the well-measured $griz$ bands. Among the galaxies observed, SDSS 031749.30+011337.1 and CGCG 432-030 have AGN as 
classified by \cite{blueE} using
emission-line ratios from the SDSS spectra. \cite{blueECO} reported a CO detection of SDSS 111850.04+422541.8, with a double-peaked 
disk-like profile and implied total molecular-gas mass near $6 \times 10^8 $ M$_\odot$. These small-scale spiral patterns are similar
to those sometimes seen in recent major mergers, such as NGC 7252 \citep{Whitmore1993}, NGC 3256 (as shown in the figures by
\citealt{Mulia}), and even 2MASX J01392400+2924067 seen before coalescence of the nuclei \citep{Koss}, which could support the conjecture of
\cite{blueECO} that such mergers are one route to producing blue galaxies with elliptical-like properties. The galaxies in our sample
share the radial scales of central spiral patterns with the nearby post-merger systems, as traced both by star clusters and by dust lanes.
The dust spirals have radial extents 1.2--3.3 kpc, and the patterns traced by bright star-forming regions span 1.2--2.6 kpc. These
are comparable to the values 1.8--6.4 kpc (dust) and 1.2--3.0 kpc (star clusters) seen in nearby merging and post-merger systems. The smaller
values apply to NGC 7252, which is the oldest local merger based on comparison with simulations and ages of star clusters, and thus
more comparable to our systems where merger signatures in the starlight distribution must be even more subtle.

There is clearly more to be done in defining how these blue early-type galaxies relate to normal ellipticals, mergers,
and even rejuvenated spirals. We plan to consolidate these HST images along with the new deep ground-based surveys to
address this in future work.


\subsection{Red spirals}

\begin{figure*}
\includegraphics[width=175.mm,angle=0]{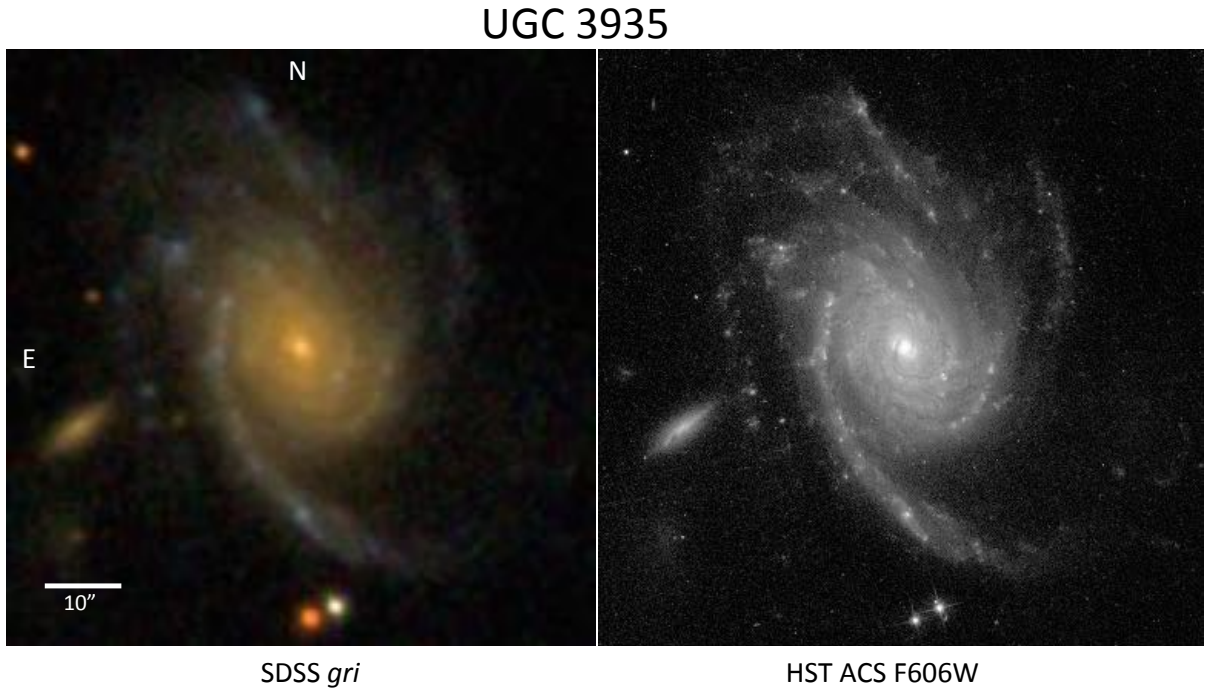}
\caption{Red spiral galaxy UGC 3935, comparing the usual $gri$ composite from SDSS data with the HST ACS F606W image.
The region shown is $77 \times 78$\arcsec with north at the top.
\label{fig-UGC3935}}
\end{figure*}

One of these, UGC 3935, has been observed. The arms include star-forming knots, while the dust arms include a
spiral pattern cutting across stellar structure near the core. This object was included in the MaNGA survey's integral-field spectroscopy 
 \citep{Bundy},
and in the associated and ongoing H I survey (\citealt{HIMANGA1}, \citealt{HIMANGA2}), which shows $3.7 \times 10^{10}$ M$_\odot$ of neutral hydrogen. This is close to the derived stellar mass $4.6 \times 10^{10}$ M$_\odot$ from SDSS data using the 
Portsmouth models \citep{PortsmouthModels}, so the galaxy's red color is not due purely to gas exhaustion. The HST image is
compared to an SDSS color-composite in Fig. \ref{fig-UGC3935}. The arms contain blue star-forming knots, and this
could be classified as a 3-armed system as well.

\subsection{Disk structures}

The range of disk structures included in Zoo Gems data is sampled in Fig. \ref{fig-diskmontage}. 

\subsubsection{Circumnuclear disks and bars}

\begin{figure*}
\includegraphics[width=175.mm,angle=0]{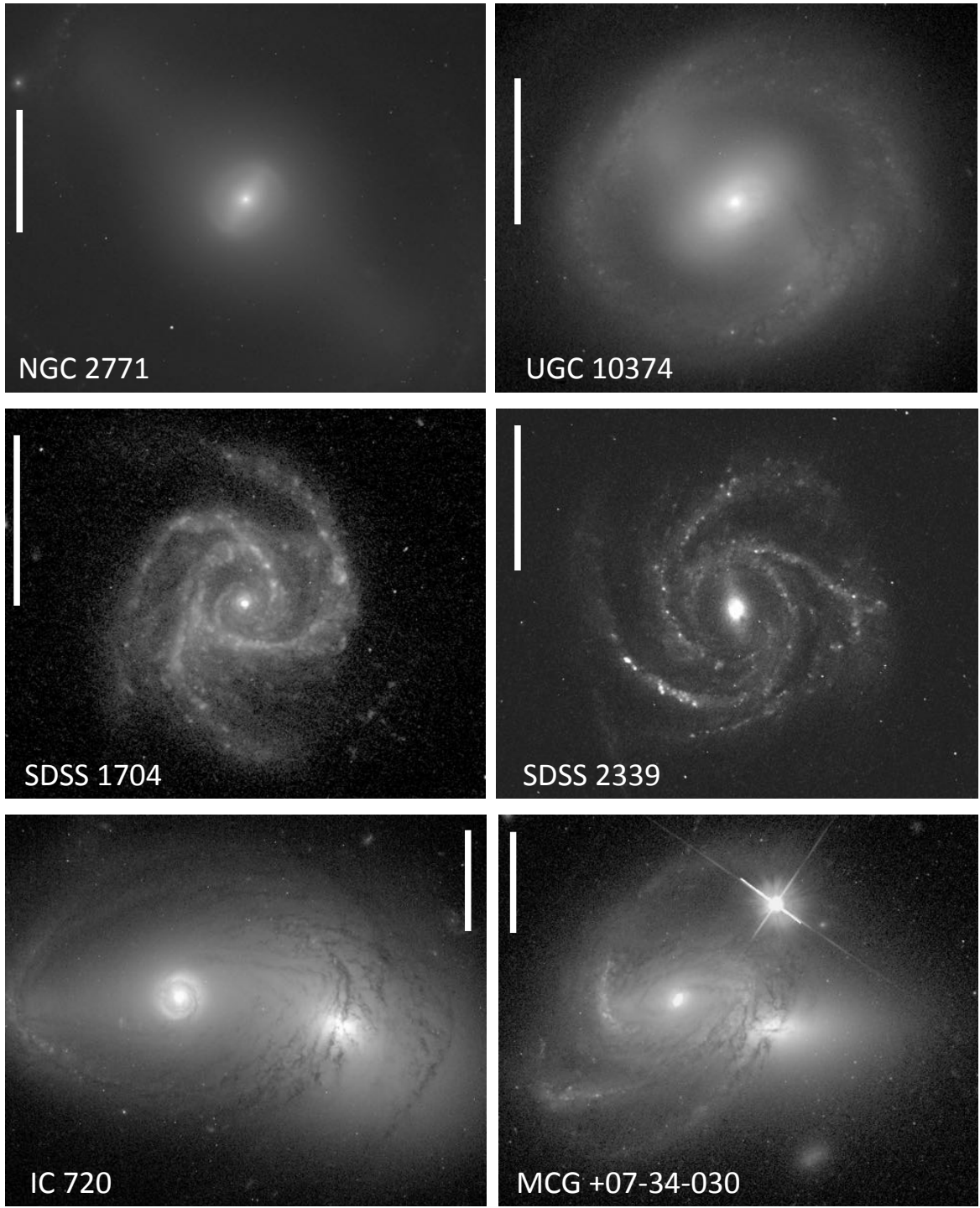}
\caption{Some of the kinds of disk structures included in Zoo Gems data. Top row, nuclear bars and barlenses. 
Middle row, 3-armed spirals. Bottom row, backlit spiral arms and disk with dust attenuation. All images have north at 
the top and east to the left; vertical white scale bars indicate 10\arcsec\  in each case. The gray scale is logarithmic, with
zero levels and contrast tailored to show the structures in each object. In the object names, SDSS destinations are
truncated for legibility.
\label{fig-diskmontage}}
\end{figure*}

The two systems shown in Fig. \ref{fig-diskmontage} both have nuclear bars and surrounding rings or barlenses (as
defined by \citealt{barlens}), which
extend beyond the bar width in each case.
The inner region of NGC 2771 forms a striking echo, rotated nearly 90$^\circ$, of the outer bar and ring of the galaxy disk.
NGC 2595 from the "unusual spiral patterns" category
shows similar features.
UGC 10374 has an outer pseudoring beyond the area shown. The backlit spiral in IC 720 shows a nuclear spiral and possibly nuclear bar which were
not well resolved even in subarcsecond ground-based images.

\subsubsection{Three-armed spirals}

Details in the arm structures of these systems may help understand why these do not show a preference for low-density
environments, as might be expected from a straightforward analysis of spiral modes after perturbation (\citealt{Elmegreen1992},
 \citealt{Hancock}). 
In SDSS J170414.33+620234.0, among the first targets observed in this program, the enhanced angular resolution of
HST images reveals quite different distributions of star-forming knots along each arm, and
the arms starting from an off-center ring around the core.

\subsubsection{Backlit dust}

Fig. \ref{fig-diskmontage} shows two of the most striking backlit-galaxy systems observed in Zoo Gems.
While we will present a full modeling of these systems elsewhere, it is already noteworthy that each case shows at least thin
arms of attenuation farther out than the detected starlight, and nearly transparent regions between the dust arms (and
within the resonance ring in the case of IC 720). The outer dust lanes illustrate in a vivid way one advantage of dust detection with this method - arbitrarily cold dust is detectable, unlike direct far-IR measurements which rely on reradiation of absorbed starlight \citep{Domingue}. These images also illustrate the gain in understanding structure
on going from SDSS to HST angular resolution. As has been shown with WFPC2 images of two backlit spirals \citep{thickthin3}, 
better linear resolution reduces the confusion between unresolved dust structure and sleep of the reddening law, as there is less blending of areas with quite different attenuation. The Zoo Gems overlap systems provide a significant addition to the range of disk structures and 
backlit regions observed with Hubble's resolution.

\subsection{Ring features}\label{ring1330}

\begin{figure*}
\includegraphics[width=175.mm,angle=0]{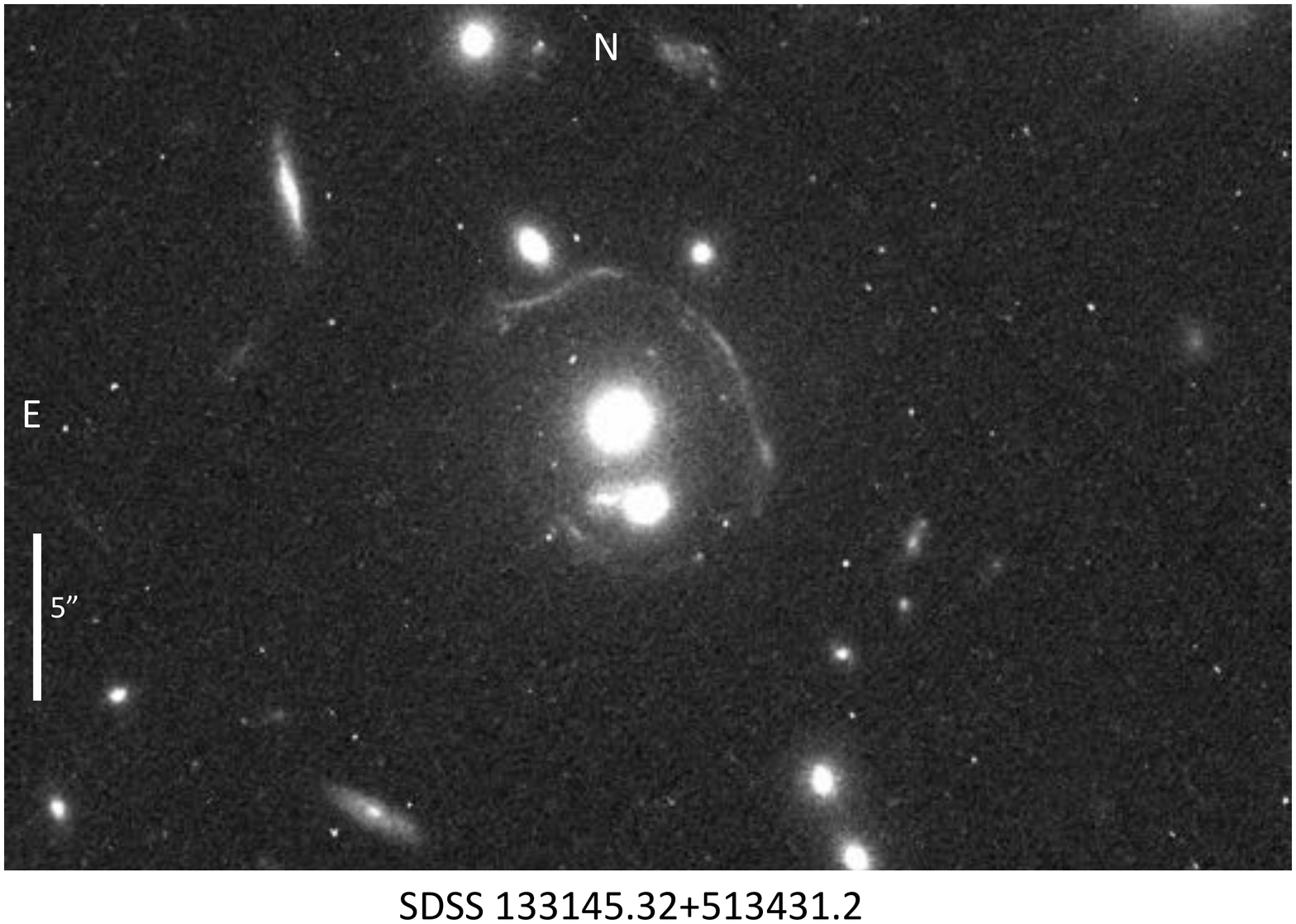}
\caption{Deep-red F814W image of the lensed arc around the central galaxy SDSS 133145.32+513431.2 ($z=0.2894$). The upward inflection to the north may indicate perturbation by the separate potential of the bright galaxy just outside the arc. The partial ring has
characteristic radius 4.0\arcsec, projecting to 17.5 kpc at the distance of the lensing system.
\label{fig-ring1331}}
\end{figure*}

Among these, SDSS 133145.32+513431.2 is especially noteworthy. This object shows a subtle annular color pattern in SDSS
and, more clearly, DES data
\citep{DESDR1}. The ACS image reveals a partial Einstein ring (Fig. \ref{fig-ring1331}), with typical radius 4.0\arcsec , and two main segments together spanning nearly 300$^\circ$ around the brightest galaxy in a group
(SDSS data give $z=0.2894$ for this galaxy). An inflection at its northern end is associated with an individual luminous galaxy. 
While the
source redshift remain unknown, pending further information on the lensed source and additional foreground group members, knowing the lens redshift we can bound the group mass enclosed within the Einstein ring (radius 4.0\arcsec or 17.5 kpc). If the source-lens distance is 
equal to the lens-observer distance (0.90 Gpc), the mass within 17.5 kpc would be $9 \times 10^{12}$ solar masses, dropping 
to $3 \times 10^{12}$ for the unrealistic case of arbitrarily high redshift.

\begin{figure*}
\includegraphics[width=175.mm,angle=0]{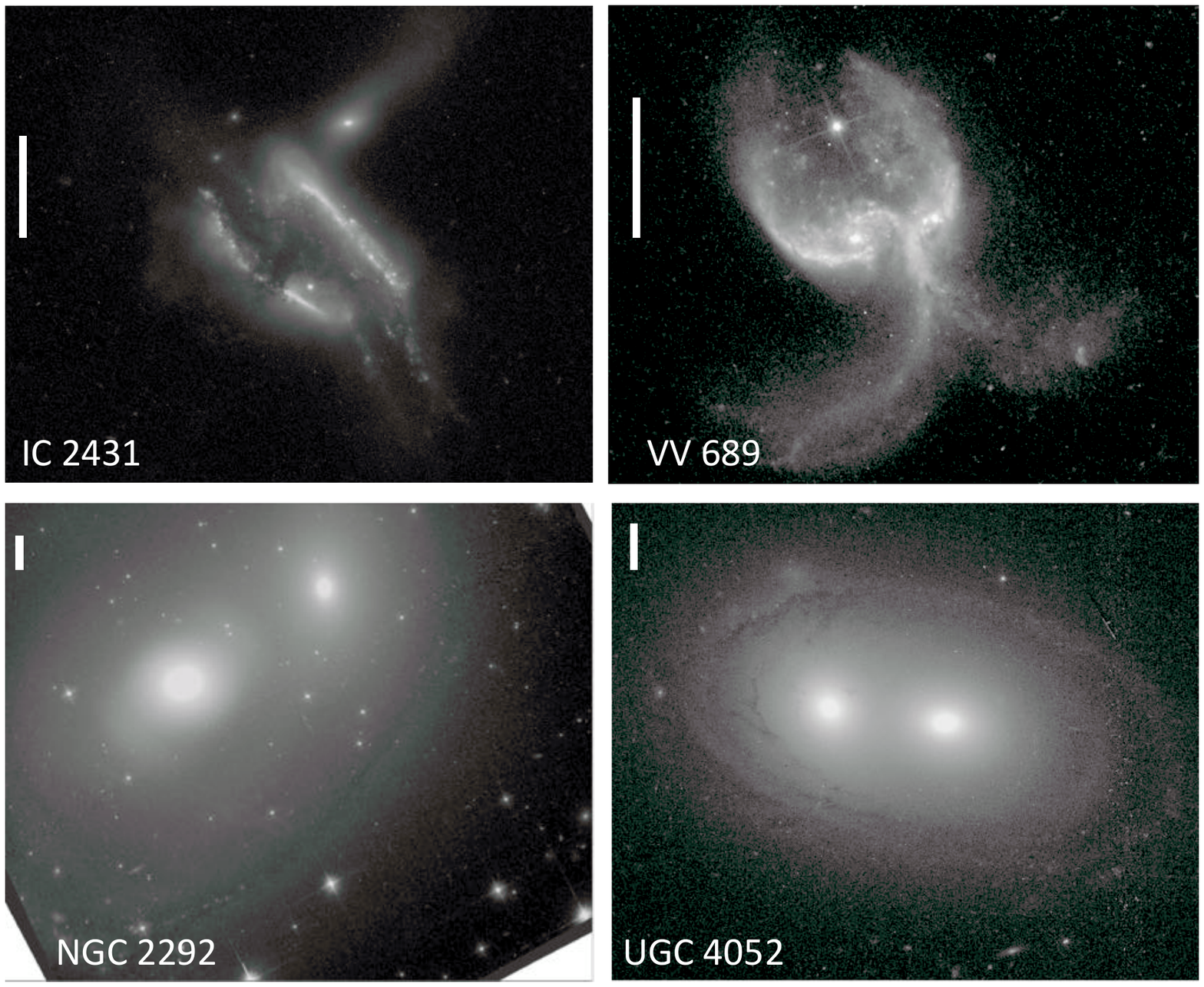}
\caption{Montage of merging galaxies, including two typical mergers (upper row) and
two potential regrowing disks (lower row). The layout is as in
Fig. \ref{fig-diskmontage}, with 10\arcsec scale bars. The regrowing disks illustrated here span large enough areas on the sky to include the
stripe of cosmic-ray events across the middle, where only a single exposure was obtained as the telescope made
a dither motion to give double coverage elsewhere. For NGC 2292, the edges of the ACS combined field appear in the corners.
\label{fig-mergermontage}}
\end{figure*}


\subsection{Regrowing disks}

The brightest of these are shown in Fig. \ref{fig-mergermontage}, in comparison to two more typical advanced merging systems.
The images illustrate the characteristics of two bulges, jointly surrounded by a disk with incomplete spiral patterns containing
dust lanes and patchy star-forming regions.

After program submission, one of these objects (CFHTLS1220-215555)  was seen to have a star superimposed on a single bulge as seen in Legacy Survey
images \citep{Dey}, which were not clearly separated in SDSS data and appeared as a double bulge. This object does
have the unusual combination of a dominant, off-center ring without a bar, and short spiral features.

The overlapping-galaxy system UGC 7064A may more exactly belong with these systems, on examination of the dust geometry.
The eastern bulge component is surrounded by a ring including stars and dust, which may also encircle the inclined
spiral disk to its west.

\subsection{Reassignment of galaxy categories}
Some objects turn out to be something quite different than we inferred from the SDSS or Legacy Survey images. For example,
the components of the NGC 5021 system show obvious signs of tidal interaction (off-center nucleus, helical dust lanes, spokes), 
rather than being a superposition of relatively undisturbed galaxies. SDSS-1237668504364187727 (SDSS-J 161224.60+594610.4), observed as a potential
reddened AGN with bluer outflow regions, looks like a multicomponent merger; a Lick Observatory spectrum, using the
Kast spectrograph at the 3m Shane telescope,
shows the core to host highly reddened star formation, with relatively unreddened star formation on either side. The
central region has emission-line redshift $z=0.1019$.

\section{Conclusions}\label{sec:conclusions}

We have described the diverse scientific cases addressed in the ``Gems of the Galaxy Zoos" HST gap-filler program, and detailed
the target selection and public input involved in its object list. The results so far illustrate the value of even short-exposure images
at Hubble's high angular resolution, deriving additional results from the effort invested by volunteers in the Galaxy Zoo and Radio Galaxy Zoo projects. These data have revealed extended redder components around Green Pea starburst systems, 
small-scale spiral patterns in blue early-type galaxies, and suggested that some merging galaxies quickly reform star-forming disks.
Additional uses will certainly be forthcoming.
The variety of results already found from this unusually wide-ranging program may encourage the community to consider ways to
achieve a similar richness of use for projects on other facilities. This includes availability of high angular resolution,
which has proven crucial in
such applications as substructure of Green Pea starbursts, using dust attenuation to unravel the geometry of disks and
merging systems, and distinguishing key morphological features such as spiral arms in radio-AGN hosts. Many of our results
illustrate the complementary roles of deep survey images and even shallow high-resolution images.



\acknowledgments
This work was enabled by the many volunteer participants in Galaxy Zoo, and especially by the beta testers and
commentators on the voting interface, those who pointed to interesting objects discussed in the project Forum and Talk
sites, and all who voted on target selection. We particularly note contributions by Ivan Terentiev, Chris Molloy, Victor Linares,
Alexander Jonkeren, Christine MacMillan, Richard Nowell, Graham Mitchell, Claude Cornen, and Michael Peck.
At STScI, program coordinator Blair Porterfield gave tips which improved the quality of our data, while John Mackenty 
was helpful in scheduling questions and in tracking down changes in scheduling priority during the program.
We thank Julianne Dalcanton for conversations on observation setup and on coordinating object lists between two
gap-filler projects.  A timely Excel suggestion from Nathan Keel greatly speeded the production of Table 1. 

Funding for the SDSS and SDSS-II has been provided by the Alfred P. Sloan Foundation, the Participating Institutions, the National Science Foundation, the U.S. Department of Energy, the National Aeronautics and Space Administration, the Japanese Monbukagakusho, the Max Planck Society, and the Higher Education Funding Council for England. The SDSS Web Site is http://www.sdss.org/.

The SDSS is managed by the Astrophysical Research Consortium for the Participating Institutions. The Participating Institutions are the American Museum of Natural History, Astrophysical Institute Potsdam, University of Basel, University of Cambridge, Case Western Reserve University, University of Chicago, Drexel University, Fermilab, the Institute for Advanced Study, the Japan Participation Group, Johns Hopkins University, the Joint Institute for Nuclear Astrophysics, the Kavli Institute for Particle Astrophysics and Cosmology, the Korean Scientist Group, the Chinese Academy of Sciences (LAMOST), Los Alamos National Laboratory, the Max-Planck-Institute for Astronomy (MPIA), the Max-Planck-Institute for Astrophysics (MPA), New Mexico State University, Ohio State University, University of Pittsburgh, University of Portsmouth, Princeton University, the United States Naval Observatory, and the University of Washington.
%

\vspace{5mm}
\facilities{HST(ACS), Lick(Shane)}







\end{document}